\documentstyle[12pt]{article}
\date{}

\begin{document}

\centerline{\LARGE Pionic Fluctuations of Constituent Quarks}
\medskip
\centerline{\LARGE and the Neutron Charge Radius}
\bigskip
\bigskip
\centerline{\large L. Ya. Glozman$^1$ and D. O. Riska$^2$}
\bigskip
\bigskip

\centerline{\it $^1$ KEK, Tanashi Branch, Tanashi, Tokyo 188-8501, Japan }
\centerline{\it $^2$Department of Physics, P.O.B. 9, University of
Helsinki, 00014 Finland}

\vspace{1cm}

\centerline{\bf Abstract}
\vspace{0.5cm}

Pion loop fluctuations of the constituent $u$ and $d$ quarks 
are shown to give only a minute contribution to 
the intrinsic charge radius of the
neutron, under the assumption that the pion-quark coupling
constant has its conventional value, with a cut--off scale
of the order of 1.0 GeV. The contribution from 
the pion loops to the anomalous 
magnetic moment of the neutron represents a small
($\sim$ 12-14\%) increase over
of the static quark
model value.
\bigskip
\bigskip
\bigskip

The bulk of the empirical value $-0.117 \pm 0.002$ fm$^2$ \cite{Koester} 
for the mean square radius of the neutron 
is accounted for by its magnetic moment alone
\cite{Weise}. The mean square radius of the neutron is
defined as \cite{Yennie}:
$$<r_n^2>= -6\lim_{q^2\rightarrow 0}{d\over
dq^2}G_E(q^2),\eqno(1)$$
where $G_E$ is the electric form factor. 
In terms of the Dirac and Pauli form factors
$G_E=F_1-(q^2/4m_n^2)F_2$, where $F_1(0)=0$ and
$F_2(0)=-1.91$ for the neutron, insertion in (1)
yields the expression
$$<r^2_n>= -6\lim_{q^2\rightarrow 0}{d\over
dq^2}{F_1(q^2)}
+{3\over 2}{F_2(0)\over m_n^2},
\eqno(2)$$
where $m_n$ the neutron mass. 
The numerical value of the latter (magnetic moment) term
is --0.126 fm$^2$, and thus is already very close to the
the empirical value. 
The (very) small difference between this value, and the
empirical value, may be viewed as an intrinsic mean
square charge radius of the neutron \cite{Weise}:
$$<r^2>_{int}=<r^2>_{exp}-<r^2>_{mag}=+0.009\pm 0.002 fm^2.\eqno(3)$$
This small positive value has to arise from the $q$-dependence
of the 
Dirac form factor $F_{1}(q^2)$. 

We here show that the lowest order pionic loop fluctuations 
of the constituent quarks (Fig. 1) lead to only a very
small  contributions
to the term of order $q^2$ in $F_{1n}(q^2)$, which yields a
contribution to $<r_n^2>_{int}$, the magnitude of which with
conventional parameter choices are even smaller than this small
empirical value.

The pion coupling to the $u$ and $d$ constituent quarks will be
assumed to have the form
$${\cal L}_{\pi qq}=i{f_{\pi qq}\over m_\pi}\bar \psi
\gamma_5\gamma_\mu\vec \tau\cdot
\partial_\mu \vec \phi \psi.\eqno(4)$$
Here the pion-quark coupling constant is determined by the $\pi NN$
coupling constant $f_{\pi NN}\simeq 1$ as $f_{\pi qq}\simeq 3/5 f_{\pi
NN}$. As the pions should decouple from the constituent quarks above
the chiral symmetry restoration scale $\Lambda_\chi\sim 4\pi f_\pi\sim
1$ GeV, a corresponding high momentum cut-off factor will be
introduced at the pion quark vertices.

The calculation of the pionic loop fluctuations to nucleon
properties may be justified at the level of constituent quarks
when -- as in the present case -- the smallness of the effective
pion-quark quark coupling suggests a converging loop expansion.
Corresponding attempts to calculate
such contributions as pion-nucleon fluctuations fail to
yield realistic results because of the much larger
pion-nucleon coupling constants and large number of
intermediate baryon resonances that have to be
considered in the loops in the hadronic approach 
\cite{Houriet,Bethe}.
At the quark level the loop contributions 
automatically take into account 
all intermediate baryon states.

The pionic fluctuations illustrated by the Feynman diagrams
in Fig. 1 both contribute to the $q-$dependent terms in the
Dirac form factors of the $u-$ and $d-$quarks. Denoting 
those $F_{1u}$ and $F_{1d}$ respectively, when
normalized to unity at $q^2=0$, the $SU(6)$
wave function for the nucleon yields the following
expression for the
intrinsic charge radius of the neutron:
$$<r_n^2>=-4\lim_{q^2\rightarrow 0}{d\over dq^2}
\{F_{1u}(q^2)-F_{1d}(q^2)\}.\eqno(6)$$

\begin{figure}[ht!]
\input FEYNMAN
\textheight 400pt \textwidth 400pt

\bigphotons
\begin{picture}(25000,25000)
\drawline\fermion[\E\REG](0,0)[13500]
\drawline\scalar[\NW\REG](11000,0)[3]
\drawline\photon[\N\LONGPHOTON](6650,4200)[5]
\drawline\scalar[\NE\REG](2500,0)[3]
\put(6600,-6000){a}
\put(2250,1750){$\pi$}
\put(10500,1750){$\pi$}

\drawline\fermion[\E\REG](18500,0)[2500]
\drawline\fermion[\E\REG](29000,0)[2500]
\drawline\fermion[\NE\REG](20800,0)[5850]
\drawline\fermion[\NW\REG](29200,0)[5850]

\drawline\scalar[\E\REG](21000,0)[4]
\drawline\photon[\N\LONGPHOTON](25000,4200)[5]
\put(25000,-6000){b}
\put(25000,1000){$\pi$}

\end{picture}
\caption{Pionic fluctuations of the constituent quarks. The
fermion lines represent $u$ and $d$ quarks. Form factors are
included at the vertices as explained in the text.}
\end{figure}
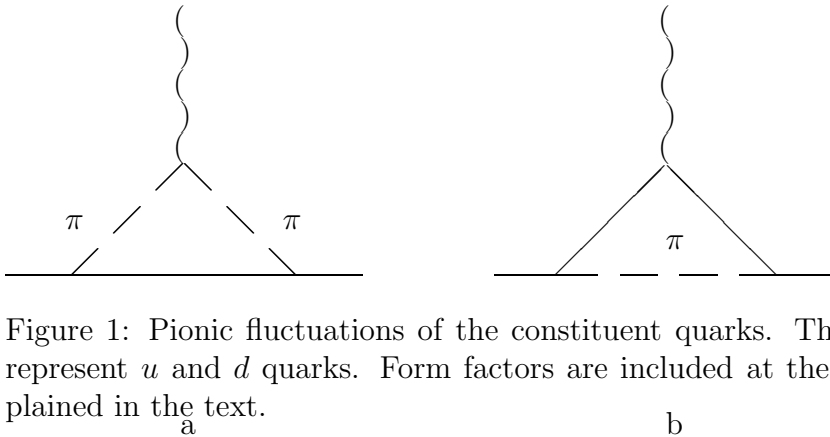

\vspace{2cm}

Here we calculate the contributions to the Dirac form factors
$F_{1u}$ and $F_{1d}$ from all the pion loop amplitudes 
that may be represented by the Feynman diagrams in
Fig. 1, where the fermion lines represent $u$ and $d$
quarks, and the pion lines represent all the pion
charge states that are allowed by charge conservation.
Two corresponding seagull diagrams are also
generated by the point
couplings
$${\cal L}_{\pi qq\gamma}=ie{f_{\pi qq}\over m_\pi}
\bar\psi\gamma_5\gamma_\mu(\vec \phi\times\vec\tau)_3\psi,
\eqno(7)$$
which arise by minimal substitution of the e.m. vector
potential $A_\mu$ in the derivative coupling (4).
Inclusion of the seagull diagrams is required by
current conservation.

In the absence of hadronic vertex form factors the
seagull diagrams may be dropped, if in place of the
derivative coupling (4) the pseudoscalar coupling:
$${\cal L}_{\pi qq}=i g_{\pi qq}\bar\psi\gamma_5
\tau\cdot\phi\psi, \eqno(8)$$
is employed. The result obtained with this
pseudoscalar pion-quark coupling is then equivalent to
that obtained with the pseudovector coupling with
inclusion of the seagull terms for the diagrams
considered here.

This equivalence may be retained also in the presence
of hadronic vertex form factors, provided that these are
included in a way that maintains current conservation.
Denote the vertex form factor to be inserted at each
pion-quark vertex in the amplitudes, where the e.m. field
couples to the quarks (Fig. 1b) $f(k^2)$ ($f(-m_\pi^2)=1$).
In these amplitudes the two form factors may be combined
with the pion propagator to a ``modified'' pion
propagator
$$v(k^2)={1\over m_\pi^2+k^2}f^2(k^2).\eqno(9)$$
We shall here take the form factor function $f(k^2)$
to have the monopole form $(\Lambda^2-m_\pi^2)/
(\Lambda^2+k^2)$, where the value of the 
parameter $\Lambda$ is $\sim \Lambda_\chi$.

To maintain current conservation, and the equivalence
between the pseudovector (including seagull terms) 
and pseudoscalar coupling models, the vertex factors
have to be inserted into the amplitude that corresponds to
the diagram Fig.1a, where the e.m. field couples to the
pion current, so that the product of the two pion
propagators is modified to \cite{Gross,Surya}:
$${1\over m_\pi^2+k_2^2}{1\over m_\pi^2+k_1^2}\rightarrow
{v(k_1^2)-v(k_2^2)\over k_2^2-k_1^2}.\eqno(10)$$
Here $k_1$ and $k_2$ represent the 4-momenta of the
pion before and after the electromagnetic coupling.
Form factors inserted by this method allows the
calculation to proceed on the basis of the pseudoscalar
coupling (8) without further reference to seagulls.

The e.m. vertex of the internal constituent quarks, 
which should be assumed to have a spatially
extended structure that may be described by a form
factor $F_q(q^2)$, is described by the current matrix element
$$<p'|j_\mu(u,d)|p>=ie({2\over 3},-{1\over 3})
\bar u(p')\{\gamma_\mu+(F_q(q^2)-1)
[\gamma_\mu-{\gamma\cdot q q_\mu\over q^2}]\} 
u(p).\eqno(11)$$
This vertex, in which the
first term corresponds to pointlike quarks, maintains the
requirement of current conservation, as the form factor 
modification appears only in a purely transverse term
\cite{Gross,Coester}.

The pion should similarly be described as spatially
extended, in view of its empirically large mean
square radius. This is well described by the $\rho-$meson
pole in the time-like region. This is taken into
account through inclusion of the pion form factor
$F_\pi(q^2)$ in the e.m. coupling of the pions, which is taken
to have the form
$$<k'|j_\mu(\pi^{\pm,0})|k>=ie(\pm 1,0)\{K_\mu
+(F_\pi(q^2)-1)[K_\mu-{K\cdot q q_\mu\over q^2}]\},\eqno(12)$$ 
where $K_\mu=k_\mu\,'+k_\mu$. 
In this vertex the form factor also appears only in 
a transverse term. Note as the electromagnetic vertices 
are considered at tree level, they may contain all
information of the quark structure of the pion and
the self-dressings of the pion and the constituent
quarks without double counting.

Consider the pionic fluctuations of u and d quarks
illustrated in Figs. 1a and 1b. The only processes, in
which the the e.m. field couples to the pions, are
the fluctuations $u\rightarrow \pi^+ d\rightarrow u$
and $d\rightarrow \pi^- u\rightarrow d$. The flavor
factors of these two processes are the same, and therefore
their contributions to the $u$ and $d$ quark current
matrix elements have the same magnitude, but opposite sign.

For $u$ quarks the fluctuations with e.m. coupling to the
internal quark lines are $u\rightarrow \pi^+ d
\rightarrow u$ and $u\rightarrow \pi^0 u
\rightarrow u$. In the first the coupling is proportional
to the $d$ quark charge $-e/3$, while the flavor factor
from the quark 
charge changing vertices is $(\sqrt{2})^2$. In the
second the coupling is proportional to the $u$ quark
charge $2e/3$, while the flavor factor from the quark charge
conserving vertices is 1. Multiplication of the charge and
flavor factors shows that these two
contributions cancel exactly in the case of the $u$ quark.
In the case of the $d$ quark the product of charge and
flavor factors for the corresponding fluctuations
$d\rightarrow \pi^- u
\rightarrow d$ and $d\rightarrow \pi^0 d\rightarrow d$
are on the other hand $4/3$ and $-1/3$ respectively,
which add up to 1. Consequently fluctuations with e.m.
coupling to the intermediate quark only contribute
to the current matrix element of the $d$ quark.

The evaluation of the net loop contributions to the
form factor combination $F_{1u}-F_{1d}$ required
for the neutron mean square radius (6) is particularly
convenient, as the combined current matrix elements satisfy
the continuity equation, without e.m. coupling to the external
quark legs. In the absence of form factors one 
obtains the result:
$$<r_n^2>_{loops}={g_{\pi qq}^2\over 8\pi^2}\bigg\{
-\int_0^1 dx x^3\{{1\over H(m_\pi)}-2{m_q^2 (1-x)^2
\over H(m_\pi)^2}\}$$
$$+\int_0^1dx(1-x)^3\{{2\over H(m_\pi)}+{m_q^2 (1-x)^2\over 
H(m_\pi)^2}\}
\bigg\}.\eqno(13)$$ 
Here the function $H(m)$ is defined as
$$H(m)=m_q^2(1-x)^2+m^2x.\eqno(14)$$
The first integral on the rhs of (13)
arises from fluctuations with e.m. coupling to the pions
(Fig.1a) and the second from terms with e.m. coupling to 
the quark lines (Fig.1b). The latter is finite in the
chiral limit, while the former is divergent in that
limit. That divergence also appears in other chiral
pion field theoretical models, e.g. in the Skyrme model
\cite{Adkins}.

The monopole hadronic form factors at the
pion-quark vertices introduced so as to maintain 
current conservation as described above are included
by means of the following substitutions in the
expression (13):
$${1\over H(m_\pi)}\rightarrow {1\over H(m_\pi)}
-{1\over H(\Lambda)}-x{\Lambda^2-m_\pi^2\over
H(\Lambda)^2},\eqno(15a)$$
$${1\over H(m_\pi)^2}\rightarrow {1\over H(m_\pi)^2}
-{1\over H(\Lambda)^2}-2x{\Lambda^2-m_\pi^2\over
H(\Lambda)^3}.\eqno(15b)$$

The pionic loop contribution to the mean square neutron
radius (13) is very small and sensitive to the 
constituent mass value. With a quark mass value
of $m_q=300$ MeV the numerical values are -0.00082 fm$^2$
and 0.0018 fm$^2$ for $\Lambda=$ 800 MeV and 1.0 GeV
respectively. With $m=340$ MeV the corresponding
numerical values are -0.0048 fm$^2$ with
$\Lambda=$ 800 MeV and -0.0023 fm$^2$ with $\Lambda=$ 1 GeV.
The value is positive for $\Lambda >$ 1.2 GeV, 
the asymptotic value for $\Lambda\rightarrow \infty$ being
0.013fm$^2$ with the latter constituent mass value. 
This latter value is close to the empirical range.
The smallness of the pion loop contribution to the
mean square neutron radius is a consequence of strong
cancellations between the two terms in (13).

	The additional contribution to the mean
square radius of the neutron from the finite radii
of the pion and the quarks within the loops are
$$<r_n^2>_\pi=r_\pi^2({g_{\pi qq}^2\over 8\pi^2})
\int_0^1 dx x \bigg\{ {1\over 2}\{\ln{H(\Lambda)\over H(m_\pi)}
-x{\Lambda^2-m_\pi^2\over H(\Lambda)}\}$$
$$-2m_q^2 (1-x)^2\{{1\over H(m_\pi)}-{1\over H(\Lambda)}
-x{\Lambda^2-m_\pi^2\over H(\Lambda)^2}\}\bigg\},\eqno(16a)$$
$$<r_n^2>_q=-r_q^2({g_{\pi qq}^2\over 8\pi^2})
\int_0^1 dx (1-x) \bigg\{{1\over 2}\{\ln{H(\Lambda)\over H(m_\pi)}
-x{\Lambda^2-m_\pi^2\over H(\Lambda)}\}$$
$$+m_q^2 (1-x)^2\{{1\over H(m_\pi)}-{1\over H(\Lambda)}
-x{\Lambda^2-m_\pi^2\over H(\Lambda)^2}\}\bigg\},\eqno(16b)$$ 
respectively. 

The mean square pion radius obtained from the empirical
parameters in ref.\cite{PDG} is $r_\pi^2=0.38$ fm$^2$.
Realistic dynamical models for the baryon spectrum 
\cite{Wagenbrunn} suggest
that the (flavor averaged) mean square matter radius of the
constituent $u$ and $d$ quarks 
should be about 0.13 fm$^2$ if the empirical mean
square radius of the proton is to be reached \cite{Helminen}.
With these values the
additional contribution to the mean square radius of the
neutron from the pion and (flavor averaged) constituent
quark radius is very small.
With $m_q=300$ MeV the value is 0.00086 fm$^2$
for $\Lambda=800$ MeV and 0.0020 fm$^2$ for $\Lambda$=1.0 GeV.
With $m_q=340$ MeV the corresponding values are somewhat
smaller: 0.00013 fm$^2$ 
and 0.0014 fm$^2$ for 
$\Lambda=$ 800 MeV and
$\Lambda=$ 1 GeV respectively.
The smallness of these values is again a consequence
of cancellations between the two contributions (16a) and
(16b).

Combination of the two contributions (13) and (16) to
the neutron radius then gives the following net contributions
to the mean square radius of the neutron: with the
quark mass 300 MeV and $\Lambda$ = 800 MeV the net contribution
is only 0.00004 fm$^2$, whereas with $\Lambda$ = 1 GeV it amounts
to 0.004 fm$^2$. In the case of $m_q=340$ MeV the net
contribution is
 -0.005 fm$^2$ for
$\Lambda=$ 800 MeV and -0.0010fm$^2$ for $\Lambda=$ 1 GeV.
With the smaller quark mass value the net contribution 
reaches 0.00 fm$^2$ and thus
the empirical range
0.009$\pm$0.002 fm$^2$
if the cut-off parameter $\Lambda$ is increased to 1.2 GeV.
With the larger quark mass value the empirical range
is reached only by $\Lambda=$ 1.5 GeV.
The conclusion is in any case that
the pionic loop fluctuations of
the constituent quarks imply only a very small contribution
to the intrinsic mean square radius of the neutron.
The pionic loop contribution does therefore
not perturb the satisfactory
description of the (bulk of the)
negative mean square radius of the neutron, which
is implied by the empirical value of the neutron magnetic moment.
 
The pionic fluctuations of the constituent quarks also give 
a small contribution to the magnetic moment of the neutron.
In units of nuclear magnetons this contribution is
$$F_2(0)_{loops}=-{g_{\pi qq}^2\over 12\pi^2}m_p m_q
\int_0^1 dx(1-x)^2(2+3x)\{{1\over H(m_\pi)}
-{1\over H(\Lambda)}-x{\Lambda^2-m_\pi^2\over H(\Lambda)^2}\}.
\eqno(17)$$
Here $m_p$ is the proton mass. The numerical value of this
contribution is only 
$\sim -0.23 $ n.m. for $\Lambda=800$ MeV and
$\sim -0.26 $ n.m. for $\Lambda=1$ GeV with
$m_q = 340$ MeV. 
The asymptotic value for $\Lambda\rightarrow\infty$ is 0.37 n.m.
These values represent enhancements of about $\sim$ 12-14\%
of the static quark model value
$-2/3(m_p/m_q)$. The latter agrees with the empirical
value --1.91 n.m. with
$m_q=345$ MeV, but only if the reduction caused
by the considerable
``relativistic correction'' that appears if the quarks are described 
by the Dirac current operator is neglected \cite{Close}. 
Hence there is a room for 
this loop correction along with exchange current contributions that
are associated with the interaction between the confined
quarks \cite{Dannbom,Helminen,Glozman}.

The pion loop contribution to the anomalous magnetic moment
of the proton $F_2^p(0)$ may also be calculated using the 
expression (17), provided that the bracket (2+3x) in the
integrand is replaced by the expression -(1+9x)/2. Using
the same parameter values as above for the neutron, we
find the loop contribution to
$F_2^p(0)$ to fall in the range 0.17 -- 0.19 n.m.
The static quark model value for the magnetic
moment of the proton is $m_p/m_q$, and for the ratio of
the neutron to proton magnetic moments $\mu_n/\mu_p=2/3$.
This ratio, which is slightly below the empirical value 0.68,
would increase by $\sim$ 5\% by inclusion of the loop
contributions considered here.

The pionic loop contribution to the neutron magnetic moment
was reported to be much larger than the value above in refs.
\cite{Ito1,Ito2}, without supporting formalism. The suggestion
in refs. \cite{Ito1,Ito2} that the contribution of the pionic 
(and kaonic and $\eta-$meson)
loop fluctuations of the constituent quarks 
to their anomalous magnetic moments
explain all of the
negative empirical mean square radius of the neutron 
is however not
tenable as it neglects the other (main) quark 
contributions that make
up the (empirical) anomalous magnetic moment of the neutron.
In addition the
contribution to the neutron mean square radius from the Dirac
form factors of the constituent quarks to the first term
of eq. (2) was neglected in refs. \cite{Ito1,Ito2}. 

The present finding is that the lowest order pionic loop fluctuations
are insignificantly small for for conventional
values of the coupling and cut--of scale parameters.
These loop corrections are the leading ones
in a $1/m$ expansion, being of order $(1/m)^2$. Small
additional contributions that are of order $(1/m)^3$ arise from
exchange currents \cite{Helminen, Glozman}.

The restriction to $SU(2)$ flavor symmetry above is not essential,
and the $K s$ and $\eta s$ fluctuations of the $u$ and $d$ quarks
may calculated by methods similar to those used above. The much
larger masses of the $K$ and $\eta$ mesons make the numerical
contributions of those fluctuations smaller than those of the pionic
fluctuations however. Vector meson loop fluctuations are 
expected to be
small, because of the much larger meson masses. Loop fluctuations in
which intermediate vector mesons undergo radiative decay to
pseudoscalar mesons give no contribution at all to the intrinsic
charge radius of the neutron. This is direct consequence of the
transversality of the transition current matrix elements, which
have the generic form
$$<\pi(k')|J_\mu|V_\sigma^b(k)>=i{g_{V\pi \gamma}\over m_V}
\epsilon_{\mu\lambda\nu\sigma} k_\lambda k'_\nu\delta^{ab}.
\eqno(18)$$
Here $V_\sigma^b$ denotes either the $\rho$ or the $\omega$
meson field (in the latter case the isospin index is left out).

\vspace{1cm}
\centerline{\bf Acknowledgment}
\vspace{0.5cm}
This research has been supported in part by the Academy
of Finland under contract 34081. L. Ya. G. acknowledges the
hospitality of the Institute for Theoretical Physics of
the University of T\"{u}bingen.

\end{document}